\documentclass[11pt,a4wide]{article}
\usepackage{newlfont}
\usepackage{amsmath}
\usepackage{graphics}
\usepackage{float}
\usepackage{graphicx}
\usepackage{epsfig}
\begin{document}

\title{Properties of $B_c$ Mesons and Variational Constraints on their Masses}
\author{Nosheen Akbar\thanks{e mail: nosheenakbar@ciitlahore.edu.pk,noshinakbar@yahoo.com} \\
\textit{Department of Physics, COMSATS University Islamabad, Lahore Campus } \\
{Defence Road, Lahore (54000), Pakistan}}
\date{}
\maketitle

\begin{abstract}
Spectrum, radii, radial wave functions at origin, decay constants and momentum widths for radial and orbital excited $B_c$ mesons are derived within non-relativistic quark model framework through finding numerical solution of the Schrodinger equation by shooting method. Masses of orbitally excited states are derived with a more simpler method that is developed by combining the uncertainty and variational principles. Masses of $B_c$ mesons are also calculated by using Momentum widths. Besides calculations, theoretical results are compared with the experimental observations which have implications for scalar form factors and leptonic decays of $B_c$ mesons.

 \end{abstract}

%%% ----------------------------------------------------------------------
%%% ----------------------------------------------------------------------

\section*{I. Introduction}

$B_c$ meson, with beauty(b) and charm(c) quarks, is the most important meson for understanding of Quantum Chromodynamics (QCD) due to its different flavoured heavy quark-antiquark pair. Its mass lies in between charmonium and bottomonium mesons. The ground state of $B_c$ meson is discovered in 1998 at Fermilab in Collider Detector \cite{CDF} with mass $6.2749 \pm 0.0008$ GeV. After decades of $B_c(1 S)$ discovery, ATLAS Collaboration observed a mass of $6.842 \pm 0.009 GeV$ for excited state $B_c$ meson \cite{ATLAS}, but this result is not confirmed by LHCb Collaboration \cite{LHC}. Recently, CMS Collaboration \cite{CMS} observed two peaks for the excited states of $B_c$ meson, $B^+_c(2 ^1 S_0)$ and $B^{+\ast}_c(2 ^3 S_1)$. The mass of $B^+_c(2S)$ is measured to be $6.871 \pm 0.0012 \pm 0.0008 \pm 0.0080$ GeV \cite{CMS}. A mass difference of $0.0291 \pm 0.0015 \pm 0.0007$ GeV is measured between two states \cite{CMS}. However exact mass of $B^{+\ast}_c(2S)$ is unknown. The reason of the difference between ATLAS and CMS measurements could be that the peak observed by ATLAS was the result of superposition of two states( $B^+_c(2S)$ and $B^{+\ast}_c(2S)$)\cite{CMS}.

Theoretically, $B_c$ mesons have been studied through quark potential model \cite{nosheen19, 054025, ishrat19, 9402210,9511267,9806444, 054016, 1750021, 0210381,0406228,1504.07538, QiLe2019}, QCD sum rule \cite{1993,9406339,1306.3486, 9403208}, the heavy quark effective theory \cite{9412269} and lattice QCD \cite{0409090,Lattice,0305018}.

A variety of numerical techniques are available in the literature to solve the Schrodinger equation. In this paper, Schrodinger equation with non-relativistic potential model is solved numerically for $B_c$ meson using Born Openheimer formalism and adiabatic approximation for ground, radially and orbitally excited states of $B_c$ meson. This numerical solution is used to find mass, radial wave function at origin, decay constants of pseudo scalar and vector meson, $r_{rms}$ and momentum width.

In Ref.\cite{2017}, variational principle is combined with uncertainty principle to derive the constraints to the mass of $c\overline{c}$ meson with same flavour of quark and antiquark. In this paper, work is extended for $B_c$ mesons with different flavour of quark and antiquark.

Potential model used for conventional mesons is discussed in the section II of this paper which was further used to calculate radial wave functions for the ground and radially excited state $B_c$ mesons by solving the Schr$\ddot{o}$dinger equation numerically. The expressions used to find masses, radial wave function at origin, decay constant, root mean square radii and momentum width of $B_c$ mesons are also written in section II. In section III, a more simpler technique, developed by combining uncertainty principle and variational principle, is used to calculate the mass of ground state of $B_c$ meson while the results are discussed in section IV.

 \section*{II. Conventional $B_c$ mesons}
\subsection{Schrodinger Equation}
 Properties of mesons can be derived by solving the Schr$\ddot{\text{o}}$dinger equation:
\begin{equation}
H \Psi = E \Psi \label{P23}
\end{equation}
where $\emph{H}$ is the energy operator known as Hamiltonian and $E$ is the total energy of the system. Hamiltonian can be defined as:
\begin{equation}
H = H_T + H_V \label{P24}
\end{equation}
$H_T$ is the kinetic energy part of the Hamiltonian and is defined as:
\begin{equation}
H_T = \frac{\textbf{P}^2}{2 \mu} \label{P24}
\end{equation}
with $P$ as the momentum and $\mu$ as the reduced mass of the quark-antiquark system. Potential energy part of the Hamiltonian for mesons is modelled as \cite{isgur2004}:
\begin{equation}
H_V = V(r) = H^{conf}+H^{cont}+H^{tens}+H^{s.o}, \label{Vr}
\end{equation}
Here
\begin{equation}
H^{conf} = \frac{-4\alpha _{s}}{3r} + b r ,
\end{equation}
\begin{equation}
H^{cont} = \frac{32\pi \alpha_s}{9 m_q m_{\overline{q}}} (\frac{\sigma}{\sqrt{\pi}})^3 e^{-\sigma ^{2}r^{2}} \textbf{S}_{q}. \textbf{S}_{\overline{q}},
\end{equation}
\begin{equation}
H^{tens} = \frac{4 \alpha _{s}}{m_q m_{\overline{q}} r^3} S_T,
\end{equation}
\begin{equation}
H^{s.o} = (\frac{\textbf{S}_{q}}{4 m_q^2}  + \frac{\textbf{S}_{\overline{q}}}{4 m_{\overline{q}}^2}).\textbf{L} (\frac{4\alpha _{s}}{3r^3}- \frac{b}{r})+ \frac{\textbf{S}_{q}+ \textbf{S}_{\overline{q}}}{2 m_q m_{\overline{q}}}.{\textbf{L}}\frac{4\alpha _{s}}{3r^3}, \label{Vr}.
\end{equation}
In $H^{conf}$, first term describes coulomb like interaction while the second one is due to linear confinement. $H^{cont}$, $H^{tens}$, and $H^{s.o}$ describe the colour contact, colour tensor, and spin orbit interactions respectively.
$\alpha _{s}$, $b$ and $S_T$ are the strong coupling constants, string tension and the tensor operator respectively. $S_T$ is defined as:
\begin{equation}
S_T=\textbf{S}_q.{\hat{r}}\textbf{S}_{\overline{q}}.{\hat{r}}-\frac{1}{3}\textbf{S}_{q}.
\textbf{S}_{\overline{q}},
\end{equation}
such that
\begin{equation}
<^{3}L_{J}\mid S_T\mid ^{3}L_{J}>=\Bigg \{
\begin{array}{c}
-\frac{1}{6(2L+3)},J=L+1 \\
+\frac{1}{6},J=L \\
-\frac{L+1}{6(2L-1)},J=L-1.
\end{array}
\end{equation}
\begin{equation}
\overrightarrow{L}.\overrightarrow{S}=[J(J+1)-L(L+1)-S(S+1)]/2,
\end{equation}

Here, $L$ is the relative orbital angular momentum of the quark-antiquark
and $S$ is the total spin angular momentum. $H^{s.o}$ and $H^{tens}$ are equal to zero~\cite{charmonia05} for $L=0$, where in the eq.(6) $\overrightarrow{S}_{q}.
\overrightarrow{S}_{\overline{q}}=\frac{S(S+1)}{2}-\frac{3}{4}$.  and $m_{q}$ is
the constituent mass of the quarks.

\subsection{Spectrum of $B_c$ Mesons}
To find the mass of $B_c$ mesons, numerical solution of the radial Schr$\ddot{\text{o}}$dinger equation:
\begin{equation}
U^{\prime \prime }(r)+2\mu (E-H_V-\frac{L(L+1)}{2\mu r^{2}})U(r)=0.
\label{P23}
\end{equation}
is found by using the shooting method. Here $U(r)=rR(r)$, product of interquark distance $r$ and the radial wave function $R(r)$. At small distance(r $\rightarrow$ 0),wave function becomes unstable due to very strong attractive potential. This problem is solved by applying mearing of position co-ordinates by using the method discussed in ref. \cite{godfrey}. The parameters ($m_b, m_c, \alpha_s, b,\sigma$) are found by fitting the meson's mass with experimentally known mass, we got the following values. $m_b= 4.733 GeV, m_c = 1.3841 GeV, \alpha_s= 0.4035$, $\sigma=1.0765 GeV$,and $b=0.18$ $\text{GeV}^2$. Mass of a $b\overline{c}$ state is obtained after the addition of constituent quark masses in the energy $E$.
Radial wave function of $^1 S_0$, $^3 S_1$, $^3 P_2$ and $^3 P_0$ are shown in Fig. (1-4) for first three radially excited states.

\subsection{Mixed States}
Mesons having same mass of quark and antiquark satisfy the parity and charge conservation laws. But mesons with different mass of quark and antiquark do not satisfy the charge conservation law. $B_{c}$ mesons with different flovoured quark-antiquark are not eigenstates of the charge conjugation. So the states with same $J$ and $P$, but with different $S$ can mix. $B_c(^1P_1)$ and $B_c(^3P_1)$ are the states with same $J$ and $P$.
\begin{eqnarray}
\left\vert P \right\rangle  &=&\cos \phi _{M}\left\vert
^{1}P_{1}\right\rangle +\sin \phi _{M}\left\vert ^{3}P_{1}\right\rangle , \\
\left\vert P^{\prime }\right\rangle  &=&-\sin \phi _{M}\left\vert
^{1}P_{1}\right\rangle +\cos \phi _{M}\left\vert ^{3}P_{1}\right\rangle ,
\end{eqnarray}%
where $\theta _{M}$ is the mixing angle.In heavy quark limit $\phi _{M}^{o}=\tan ^{-1}\left( \sqrt{\frac{L}{L+1}}\right) $. For $L=1$,
$\phi _{M}^{o}=35.3^{o}$. For $L=2$, $\phi _{M}^{o}=39.2^{o}$. Mass for mixed states $P$, $P^{\prime}$ and $D$, $D^{\prime}$ are reported in Table 1.
\begin{table}\caption{Masses of ground, radially, and orbitally excited state $B_c$ mesons . Calculated masses are rounded to 0.0001 GeV.}
\tabcolsep=4pt
\fontsize{8}{10}\selectfont
\begin{center}
\begin{tabular}{|c|c|c|c|c|c|c|c|c|}
\hline
Meson & $J^{P}$& My Calculated & Theo. Mass & Theor. Mass~ & Theo. Mass(GI)& Lattice&  Exp. Mass \\
State & & Mass & \cite{nosheen19} &\cite{QiLe2019} & \cite{isgur2004}&&\cite{pdg} \\
 & & & & Model & &&\\ \hline
 & &  \textrm{GeV} & \textrm{GeV} & \textrm{GeV} &\textrm{GeV}& &\textrm{GeV} \\ \hline
 $ (1 ^3S_1)$ & $1^{-}$ & 6.3062 & 6.314 &6.326&6.338 & $6.331 \pm 0.004 \pm 0.006$ & \\
$ (1 ^1S_0)$ & $0^{-}$ & 6.2740 & 6.274 & 6.271&6.271 & $6.276 \pm 0.003 \pm 0.006$ &$ 6.2749 \pm 0.0008$ \\ \hline
 $(2 ^3S_1)$& $1^{-}$ &6.8845 & 6.855 & 6.890 &6.887&&\\
$ (2 ^1S_0)$ & $0^{-}$ &6.8717 & 6.841 & 6.871&6.855& &$ 6.871 \pm 0.0012 \pm 0.0008 \pm 0.0008 $\\ \hline
 $(3 ^3S_1)$& $1^{-}$ &7.2905 & 7.206 & 7.252&7.272 & &\\
 $(3 ^1S_0)$ & $0^{-}$ &7.2818 & 7.197 &7.239 &7.250 & &\\ \hline
 $(4 ^3S_1)$& $1^{-}$ & 7.6323& 7.495& 7.550 & & &\\
 $(4 ^1S_0)$ & $0^{-}$ & 7.6255 &7.488 & 7.540 & & & \\ \hline
$(5 ^3S_1)$& $1^{-}$ & 7.9374& - & 7.813 & & &\\
 $(5 ^1S_0)$ & $0^{-}$ & 7.9317 & - & 7.805& & &\\ \hline

$(1 ^3P_2)$ & $2^{+}$ &6.7362 &6.753 &6.787 & 6.768& &\\
$(1 P_1^{\prime})$ & $1^{+}$ & 6.7455 & 6.744 & 6.776 & 6.750& &\\
$(1 P_1) $ & $1^{+}$ &6.7182 & 6.7271 & 6.757 & 6.741& &\\
$(1 ^3P_0)$ & $0^{+}$ & 6.7224 & 6.701 &6.714 &6.706& &\\ \hline

$(2 ^3P_2)$ & $2^{+}$ &7.1585 & 7.111 &7.160 & 7.164& &\\
$(2 P_1^{\prime})$ & $1^{+}$ &7.1725 & 7.098 & 7.150& 7.15& &\\
$(2 P_1)$ & $1^{+}$ &7.1399 & 7.0947 & 7.134 &7.145 & &\\
$(2 ^3P_0) $ & $0^{+}$ &7.1524 & 7.086 & 7.107 &7.122& &\\ \hline

 $(3 ^3P_2)$ & $2^{+}$ & 7.5105 & 7.406 & 7.464& & &\\
 $(3 P_1^{\prime})$ & $1^{+}$ & 7.5256 & 7.393 &7.458 & & &\\
 $(3 P_1)$ & $1^{+}$ &7.493 & 7.4039 & 7.441&& &\\
$(3^3P_0) $ & $0^{+}$ &7.508 &7.398 &7.420 & & &\\ \hline

 $(4 ^3P_2)$ & $2^{+}$ & 7.8231 & - & 7.732 & & &\\
 $(4 P_1^{\prime})$ & $1^{+}$ & 7.8384 & - &7.727& && \\
 $(4 P_1)$ & $1^{+}$ &7.493 & - & 7.710& & &\\
$(4^3P_0) $ & $0^{+}$ &7.8222 &- &7.693 & & &\\ \hline

 $(1 ^3D_3)$ & $3^{-}$ & 7.0137 & 6.998 & 7.030& 7.045& &\\
 $(1 D_2^{\prime})$ & $2^{-}$ & 7.0093 & 6.984 &7.032 & 7.036& &\\
 $(1 D_2)$ & $2^{-}$ &7.0097 & 6.986& 7.024&  7.028& &\\
$(1^3 D_1) $ & $1^{-}$ &7.0008 &6.964 &7.020 & 7.041 & & \\ \hline

 $(2 ^3D_3)$ & $3^{-}$ & 7.3775 & 7.302 & 7.348&& &\\
 $(2 D_2^{\prime})$ & $2^{-}$ & 7.3755 & 7.293 &7.347& & & \\
 $(2 D_2)$ & $2^{-}$ &7.0097 & 7.294 & 7.343&& &\\
$(2^3 D_1) $ & $1^{-}$ &7.0008 & 7.280 &7.336 && & \\ \hline

\end{tabular}
\end{center}
\end{table}
\begin{table}\caption{Radial wave function at origin and decay constant of $B_c$ mesons . }
\tabcolsep=4pt
\fontsize{9}{11}\selectfont
\begin{center}
\begin{tabular}{|c|c|c|c|c|c|c|c|c|c|c|}
\hline
Meson & $J^{P}$& My Calculated &My Calculated & $|R(0)|^2$ & My Calculated &$f_p$ &$f_p$ &$f_p$ & $f_p$ & $f_p$ \\
State & & $|\psi(0)|^2$ & $|R(0)|^2$ & \cite{09735} & $f_p$  & \cite{09735} &\cite{13100941} &\cite{a} &\cite{2001}& \cite{soni}\\ \hline
 & &  $\textrm{GeV}^3$ & $\textrm{GeV}^3$& &$\textrm{GeV}$ & \textrm{GeV} & \textrm{GeV} &  \textrm{GeV}& \textrm{GeV} &\\ \hline
 $ (1 ^3S_1)$ & $1^{-}$ & $0.1557$ & 1.9566 & -& 0.5443 & 0.471&- &0.411 & 0.604 &0.435\\
$ (1 ^1S_0)$ & $0^{-}$ & 0.1782 &2.2393 &1.994 &0.5839 & 0.498 & $0.528 \pm 0.019 $&0.412 & 0.607 &0.432\\ \hline
 $(2 ^3S_1)$& $1^{-}$ &0.0956 &1.2014 &- & 0.4083 &  && & &0.356\\
$ (2 ^1S_0)$ & $0^{-}$ &0.1003 &1.2604 &1.144 &0.418 & & & & &0.355\\ \hline
 $(3 ^3S_1)$& $1^{-}$ &0.0802 &1.008 & -& 0.3634 & & & & &0.326\\
 $(3 ^1S_0)$ & $0^{-}$ & 0.0827 & 1.0392 & 0.944&0.3693 &  & & & &0.325\\ \hline
 $(4 ^3S_1)$& $1^{-}$ & 0.0727 &0.9136 & -& 0.3381 & & & & &0.308\\
 $(4 ^1S_0)$ & $0^{-}$ & 0.0745 & 0.9362& 0.8504 & 0.3423 & & & & &0.307\\ \hline
\end{tabular}
\end{center}
\end{table}

\begin{figure}[tbp]
\begin{center}
\epsfig{file=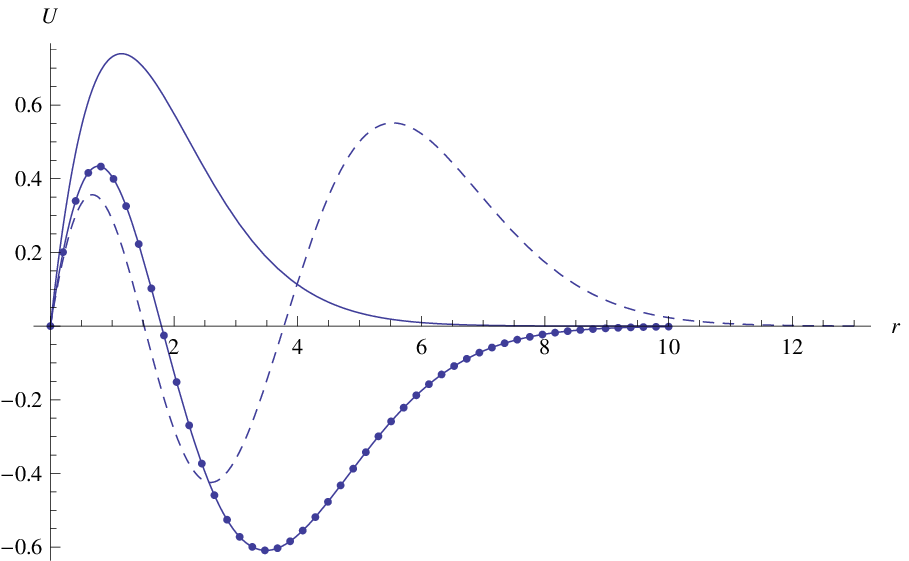,width=0.5 \linewidth,clip=}
\end{center}
\caption{$^1 S_0$ state for n=1,2,3. Line curve is for $n=1$, the curve with line plus dots is for $n=2$ and dashed curve is for $n=3$ }
\end{figure}
\begin{figure}[tbp]
\begin{center}
\epsfig{file=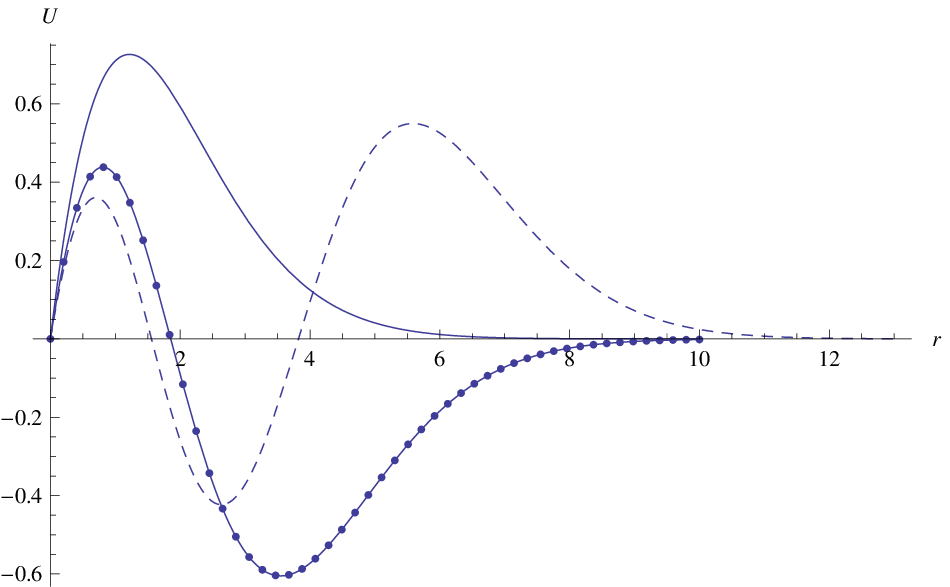,width=0.5 \linewidth,clip=}
\end{center}
\caption{$^3 S_1$ state for n=1,2,3. Line curve is for $n=1$, the curve with line plus dots is for $n=2$ and dashed curve is for $n=3$ }
\end{figure}
\begin{figure}[tbp]
\begin{center}
\epsfig{file=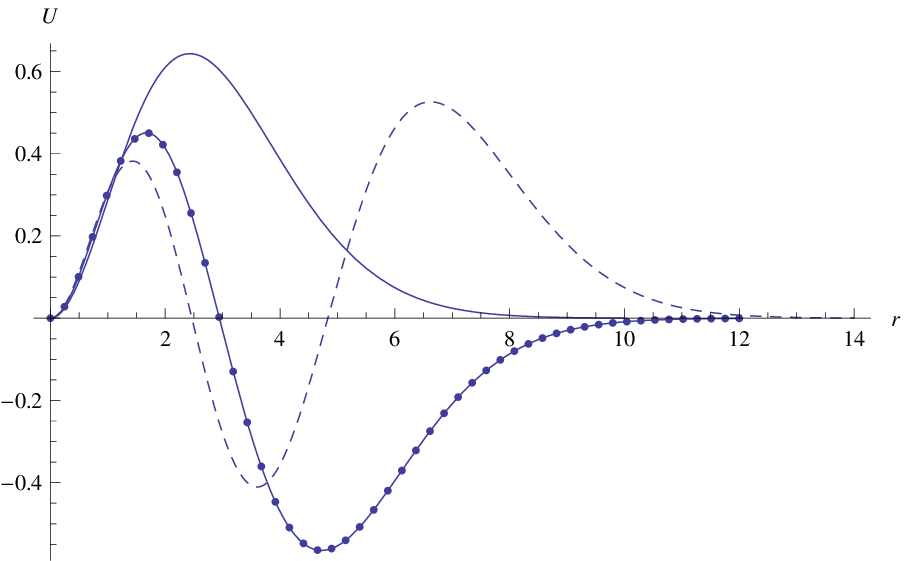,width=0.5 \linewidth,clip=}
\end{center}
\caption{$^3 P_0$ state for n=1,2,3. Line curve is for $n=1$, the curve with line plus dots is for $n=2$ and dashed curve is for $n=3$ }
\end{figure}
\begin{figure}[tbp]
\begin{center}
\epsfig{file=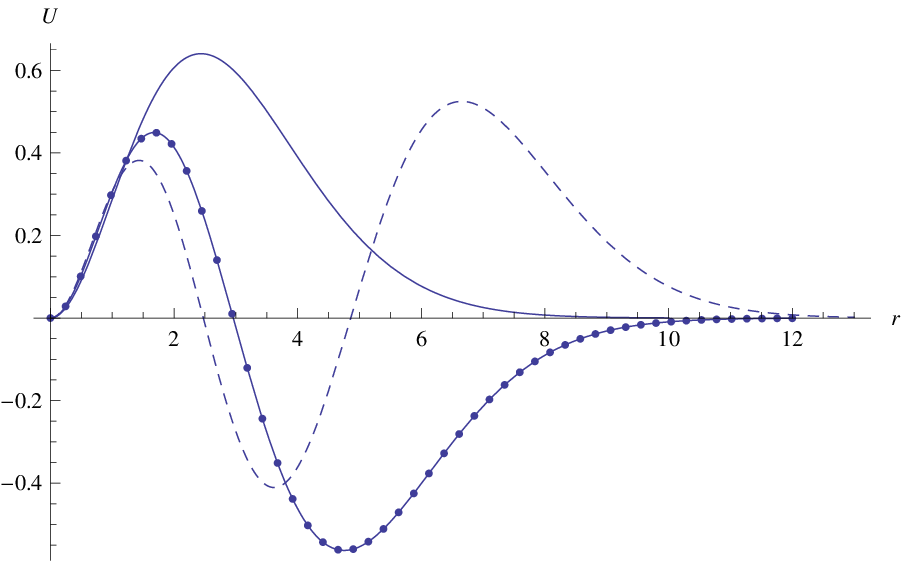,width=0.5 \linewidth,clip=}
\end{center}
\caption{$^3 P_2$ state for n=1,2,3. Line curve is for $n=1$, the curve with line plus dots is for $n=2$ and dashed curve is for $n=3$ }
\end{figure}
\subsection{Radial wave function at origin}
For normalized wave function
\begin{equation}
U^{'}(0)= R(0)= \sqrt{4 \pi} \psi(0).
\end{equation}
 $U^{'}(0)$ is calculated to find the radial wave function at origin whose magnitudes are reported in Table 2. For the states with $L>0$, wave function becomes zero at the origin.
\subsection{Decay Constants}
 $|\psi(0)|^2 $ is used
to find the decay constants ($f_p$) of pseudo scalar and pseudo vector mesons.  Following Van-Royen-Weisskopf formula \cite{VWF} is used to find decay constants.
\begin{equation}
f_p = \sqrt{\frac{3 |R(0)^2|}{\pi M_p}} =\sqrt{\frac{12 |\psi(0)^2|}{M_p}}.
\end{equation}
where $M_p$ is the mass of corresponding meson. I used the numerically calculated masses (given in Table 1) for pseudo scalar and vector meson.

\subsection{Radii }
 The normalized wave functions are then used to calculate root mean square radii using the following relation:
\begin{equation}
\sqrt{\langle r^{2}\rangle }=\sqrt{\int U^{\star }r^{2}Udr}.  \label{P25}
\end{equation}

\subsection{Momentum Width}
 Momentum width ($\beta$) for a system of quark-antiquark bound state is defined as \cite{wong04}
\begin{equation}
\beta = \sqrt{\frac{3}{2}} \frac{1}{r_{rms}}
\end{equation}
Using the root mean square radii, $\beta$ is calculated.
\begin{table}\caption{Radii and Momentum Widths of $B_c$ mesons.}
\begin{center}
\begin{tabular}{|c|c|c|c|c|}
\hline
Meson & $J^{P}$& Our Calculated & $\beta$  \\
State & & Radii & with NR potential   \\
 & & & Model \\ \hline
 & &  \textrm{fm} & \textrm{GeV} \\ \hline
 $ (1 ^3S_1)$ & $1^{-}$ & 0.345 & 0.7007  \\
$ (1 ^1S_0)$ & $0^{-}$ & 0.333 & 0.7254 \\ \hline
 $(2 ^3S_1)$& $1^{-}$ & 0.715 & 0.3378 \\
$ (2 ^1S_0)$ & $0^{-}$ &0.7079 & 0.3412\\ \hline
 $(3 ^3S_1)$& $1^{-}$ &1.0132 & 0.2384\\
 $(3 ^1S_0)$ & $0^{-}$ &1.0082 & 0.2396 \\ \hline
 $(4 ^3S_1)$& $1^{-}$ & 1.2733& 0.1897 \\
 $(4 ^1S_0)$ & $0^{-}$ & 1.2694 & 0.1903 \\ \hline

$(1 ^3P_2)$ & $2^{+}$ &0.5702 & 0.423 \\
$(1 ^3P_0)$ & $0^{+}$ & 0.5673 & 0.426 \\ \hline

$(2 ^3P_2)$ & $2^{+}$ & 0.893 & 0.2705 \\
$(2 ^3P_0)$ & $0^{+}$ & 0.8883 & 0.2719 \\ \hline

$(3 ^3P_2)$ & $2^{+}$ & 1.1656 & 0.2072 \\
$(3 ^3P_0)$ & $0^{+}$ & 1.1622 & 0.2078 \\ \hline

\end{tabular}
\end{center}
\end{table}

\section*{III. Spectrum of Mesons by Uncertainty  and Variational Principles}
Heisenberg's uncertainty principle can be written as
\begin{equation}
 \beta \overline{x} \geq \frac{1}{2}
\end{equation}
with $\Delta p_x = \beta$ (momentum width of the wave function) and $\Delta x = \overline{x}$ (size of meson corresponding to wavefunction). As Hamiltonian is the sum of kinetic and potential energy, so can be written as
\begin{equation}
H = \frac{\textbf{P}^2}{2 \mu} + H_V(r), \label{P24}
\end{equation}
or
\begin{equation}
H = \frac{p_x^2}{2 \mu} + \frac{p_y^2}{2 \mu} + \frac{p_z^2}{2 \mu} + H_V(r) \label{P24}
\end{equation}
From uncertainty principle,  $\beta \geq \frac{1}{2 \overline{x}}$. Assuming $\beta = p_x = \frac{1}{2 \overline{x}}$, Hamiltonian can be written as
\begin{equation}
H = \frac{1}{2 \mu \overline{x}} + \frac{1}{2 \mu \overline{y}} + \frac{1}{2 \mu \overline{z}} + H_V(\sqrt{x^2+y^2+z^2})  \label{P34}
\end{equation}
Assuming $\overline{x}=\overline{y}=\overline{z}$, minimize the Hamiltonean with $\overline{x}$. Using $r_{min}= \sqrt{3}x_{min}$, the $r_{min}$ is calculated. Then following expression is used to find mass of $B_c$ mesons.
\begin{equation}
M = m_Q + m_{\overline{Q}} + H  \label{P34}
\end{equation}
Combining Eq.(22)and Eq.(23),
\begin{equation}
M = m_Q + m_{\overline{Q}} + \frac{1}{8 \mu r_{min}^2} + H_V(r_{min})  \label{P34}
\end{equation}

The masses calculated by this method are reported in Table 4. For $B_c(1^1 S_0)$ state, $\%$ error is calculated by taking the experimental mass of this state, while for other states, $\%$ error is calculated by considering the masses reported in Table 1. $\%$ error reported in Table 4 show that this method is successful for ground state as well as for orbitally excited states.

\begin{table}\caption{Mass of $B_c$ mesons Using eq.(23). Calculated masses are rounded to 0.0001 GeV.}
\begin{center}
\begin{tabular}{|c|c|c|c|c|c|}
\hline
Meson  & $J^{P}$& $x_{min}$ & $r_{min}$ & Mass & \% error \\
State & &  \textrm{fm} & \textrm{fm} &\textrm{GeV} &\\ \hline
 $ (1 ^3S_1)$ & $1^{-}$ & 5.8729 &0.1128 &  0.195 & 6.87 \\
$ (1 ^1S_0)$ & $0^{-}$ & 5.8497 &0.109 &  0.189 & 6.7 \\ \hline

$(1 ^3P_2)$ & $2^{+}$ &6.6994 & 0.2355 & 0.408 & 0.55\\
$(1 ^3P_0)$ & $0^{+}$ & 6.7073 & 0.208 & 0.36 & 0.22\\ \hline

$(1 ^3D_3)$ & $3^{-}$ &7.113 & 0.341& 0.591 & 1.42 \\
$(1 ^3D_1)$ & $1^{-}$ & 7.129 & 0.331 & 0.574 & 1.83 \\ \hline

\end{tabular}
\end{center}
\end{table}

eq.(24) can be modified by replacing $r_{min}$ with $r_{rms}$. The modified form of eq.(24) can be written as

\begin{equation}
M = m_Q + m_{\overline{Q}} + \frac{\beta^2}{2 \mu} + V(r_{rms})+\frac{L(L+1)}{2 \mu r^2_{rms}}  \label{P34}
\end{equation}
The mass obtained by using $\beta$ and $r_{rms}$ in Eq.(25)are reported in Table 5.
\begin{table}\caption{Mass of $B_c$ mesons . Calculated masses are rounded to 0.0001 GeV.}
\begin{center}
\begin{tabular}{|c|c|c|c|c|}
\hline
Meson & $J^{P}$& Mass & \% error \\
State & &  by Varaitional &  \\
 &  & \textrm{GeV} &\\ \hline
 $ (1 ^3S_1)$ & $1^{-}$ &  6.3544 & 0.764 \\
$ (1 ^1S_0)$ & $0^{-}$ & 6.3478 & 0.112 \\ \hline
 $(2 ^3S_1)$& $1^{-}$  &6.6745 & 3.050\\
$ (2 ^1S_0)$ & $0^{-}$ & 6.6676 & 2.97\\ \hline
 $(3 ^3S_1)$& $1^{-}$ & 6.9635 & 4.485\\
 $(3 ^1S_0)$ & $0^{-}$  & 6.9587 & 4.437 \\ \hline
 $(4 ^3S_1)$& $1^{-}$  & 7.2126 & 5.499 \\
 $(4 ^1S_0)$ & $0^{-}$  & 7.2089 & 5.464 \\ \hline

$(1 ^3P_2)$ & $2^{+}$  & 6.6469 & 1.326\\
$(1 ^3P_0)$ & $0^{+}$ & 6.6453 & 1.147\\ \hline

$(2 ^3P_2)$ & $2^{+}$ & 6.8929 & 3.71 \\
$(2 ^3P_0)$ & $0^{+}$  & 6.8889 & 3.684 \\ \hline

$(3 ^3P_2)$ & $2^{+}$ & 7.1366 & 4.979 \\
$(3 ^3P_0)$ & $0^{+}$ & 7.1335 & 4.988 \\ \hline

\end{tabular}
\end{center}
\end{table}

\section*{IV. Discussion and conclusion}

In Fig. (1-4), radial wave functions are plotted against quark-antiquark distance. Figures illustrate that the peaks are shifted away from the origin with the orbital excitations. It is observed that the number of nodes increases by going toward higher radially excited state. In Table 1, calculated masses are reported for the ground as well as radially and orbitally excited states of $B_c$ mesons in non relativistic framework along with the experimental and theoretical predictions of the other's works. Calculated masses are in complete agreement with the already calculated theoretical masses as well as the experimental values. The results reported in Table 2 illustrate that wave function at origin and decay constants are decreasing toward higher radial excitations. Pseudo scalar $B_c$ mesons have higher values of $|\psi(0)|^2$ and $f_p$ as compare to vector mesons. The wave function at origin and decay constants are compared with others work in Table 2. In Table 3, radii, momentum widths are reported in 3rd and 4th column. It is noted that radii of $B_C$ mesons increase with radial and angular excitations. Mass calculated by using the radii and momentum widths are reported in Table 5.The $\% $ error between this mass and numerically calculated mass (reported in Table 1) shows a good agreement.
It is observed that the different states of $B_c $ mesons described with momentum width depends on $L$ and $r_{rms}$. It is concluded that the constraints derived between mass and radius by combining the uncertainty and variational principles give accurate results for ground state as well as radially and orbitally excited states.

\section{Acknowledgement}
NA acknowledge the financial support of Higher Education Commission of Pakistan through NRPU project number 7969.

\end{document}